\documentclass[iop]{emulateapj}
\submitted{ApJ, accepted 23/11/2012}
\usepackage{graphicx}
\usepackage{natbib}
\usepackage{color}

\def\ltsima{$\; \buildrel < \over \sim \;$}
\def\simlt{\lower.5ex\hbox{\ltsima}}

\def\lesssim{\mathrel{\hbox{\rlap{\hbox{\lower4pt\hbox{$\sim$}}}\hbox{$<$}}}}
\def\gtrsim{\mathrel{\hbox{\rlap{\hbox{\lower4pt\hbox{$\sim$}}}\hbox{$>$}}}}

\shorttitle{Feedback from HMXBs on the IGM: Model Spectra}
\shortauthors{Power et al.}

\begin{document}

\title{Feedback from High-Mass X-Ray Binaries on the High Redshift Intergalactic Medium : Model Spectra}

\author{Chris Power\altaffilmark{1,4}, Gillian James\altaffilmark{2}, 
  Celine Combet\altaffilmark{3} \& Graham Wynn\altaffilmark{2}}
\affil{
  $^1$ International Centre for Radio Astronomy Research, University 
  of Western Australia, 35 Stirling Highway, Crawley, WA 6009, Australia\\
  $^2$ Department of Physics \& Astronomy, University of Leicester, 
  Leicester, LE1 7RH, UK\\
  $^3$ Laboratoire de Physique Subatomique et de Cosmologie, 
  Universit\'e Joseph Fourier Grenoble 1/CNRS/IN2P3/INPG,
  53 avenue des Martyrs, 38026 Grenoble, France\\
  $^4$ ARC Centre of Excellence for All-Sky Astrophysics (CAASTRO)}
\email{chris.power@icrar.org}

\begin{abstract}

Massive stars at redshifts $z \gtrsim 6$ are predicted to have played
a pivotal role in cosmological reionization as luminous sources of 
ultra-violet (UV) photons. However, the remnants of these massive 
stars could be equally important as X-ray luminous ($L_X\!\sim\!
10^{38}$~erg~s$^{-1}$) high-mass X-ray binaries (HMXBs). Because 
the absorption cross section of neutral hydrogen decreases sharply 
with photon energy ($\sigma \! \propto \! E^{-3}$), X-rays can 
escape more freely than UV photons from the star-forming regions in 
which they are produced, allowing HMXBs to make a potentially 
significant contribution to the ionizing X-ray background during 
reionization. In this paper, we explore the ionizing power of HMXBs
at redshifts $z \gtrsim 6$ using a Monte Carlo model for a coeval 
stellar population of main sequence stars and HMXBs. Using 
the archetypal Galactic HMXB Cygnus X-1 as our template, we propose a composite
HMXB spectral energy distribution consisting of black-body and 
power-law components, whose contributions depend on the accretion state 
of the system. We determine the time-dependent ionizing power of a combined 
population of UV-luminous stars and X-ray luminous HMXBs, and deduce fitting 
formulae for the boost in the population's ionizing power arising from HMXBs; 
these fits allow for simple implementation of HMXB feedback in numerical 
simulations. Based on this analysis, we estimate the contribution of high redshift 
HMXBs to the present-day soft X-ray background, and we show that it is
a factor of $\sim 100-1000$ smaller than the observed limit. Finally, we 
discuss the implications of our results for the role of HMXBs in reionization 
and in high redshift galaxy formation.

\end{abstract}


\section{Introduction}
\label{sec:intro}

There is strong and compelling observational evidence that the Universe 
underwent an ``Epoch of Reionization'' within the first $\sim$1 billion 
years after the Big Bang 
\citep[e.g.][]{2010ApJ...723..869O,2010MNRAS.407.1328M,2011arXiv1108.3334S,2011MNRAS.415.3237M}. 
During this period, the cosmic abundance of neutral hydrogen was 
``re-ionized'' by a background of ionizing ultra-violet (UV) and X-ray 
radiation produced by the first generation of stars and galaxies 
\citep[e.g.][]{2007RPPh...70..627B,2010Natur.468...49R}. The precise 
nature of these sources remains an outstanding problem, but it can be 
argued reasonably that massive stars ($\rm M_{\ast} \gtrsim 
8\,M_{\odot}$) must have been important 
\citep[e.g.][]{2008ApJ...684....1W,2012arXiv1201.4820W}.

Massive stars have fleeting main sequence (MS) lives ($\sim$10 Myrs), 
but as extremely luminous sources of hydrogen-ionizing UV photons 
\citep[see, for example,][]{2003A&A...397..527S} they are expected to 
play a crucial role in reionizing the Universe 
\citep[e.g.][]{2003ApJ...588L..69W,2004MNRAS.350...47S,2008ApJ...684....1W}. 
However, their predicted contribution to reionization depends on the 
ease with which UV photons can escape into the intergalactic medium 
(IGM), which depends on various factors, principally the clumpiness of 
the IGM \citep[e.g.][]{2009MNRAS.394.1812P}. However, testing such 
predictions observationally is inherently challenging and a clear 
consensus as to what the sources of reionization are is yet to emerge. 
For example, observations of star-forming galaxies at redshifts $7 
\lesssim z \lesssim 10$ hint that implied star formation rates are 
insufficient to produce enough massive stars to reionize the Universe by 
$z \simeq 6$, without including a population of galaxies below the 
detection limits and relaxing assumptions about the escape fraction of 
ionizing photons and extrapolated star formation rates 
\citep[e.g.][]{2010MNRAS.409..855B,2010MNRAS.403..960M,2010ApJ...719.1250F,2010ApJ...713..115G}. 
In contrast, observations of high redshift galaxies 
also suggest that the star formation rate at $z \sim 9$ could be as high as that 
at $z$=0 \citep[cf.][]{2011arXiv1106.1745I} and that the observable population of 
galaxies between $6 \lesssim z \lesssim 8$ can sustain a fully reionized IGM at 
$z$=6 if the average escape fraction of ionizing photons is $\sim$30\% 
(cf. \citealt{2012ApJ...758...93F}; see also \citealt{2012ApJ...752L...5B}). A 
similar conclusion is drawn by \citet{2009ApJ...705L.104K}, who argue that a 
sufficient number of massive stars could have formed to reionize the Universe by 
using high-redshift gamma ray bursts as a proxy for the integrated star 
formation rate between $6 \lesssim z \lesssim 8$.\\

It is, however, worth noting that massive stars can continue to ionize 
during their post-MS lives, predominantly as sources of X-rays. 
\citet{2001ApJ...553..499O} noted that X-ray emission from supernovae in 
star forming regions in the high-redshift Universe should be large and 
comparable energetically to UV emission. The role of supernovae has been 
explored further by \citet{2011arXiv1105.3207J} who examined how strong 
shocking of the ISM associated with supernovae leads to the production 
of ionizing photons with harder spectra and larger escape fractions than 
UV photons. They estimate that such X-rays can boost -- briefly -- the ionizing 
power of a massive star by $\sim\!10\%$.

\citet{2003MNRAS.340..210G} considered not only X-rays from 
supernovae but also from X-ray binaries in their examination of the 
contribution of star formation to the build up of the high redshift X-
ray background. As \citet{2012MNRAS.tmp.2855J} have demonstrated 
in their recent study, X-ray binaries are likely to be important sources 
of feedback across cosmic time \citep[see also][]{2012arXiv1206.2395F}. 
In particular, we expect this to be the case at early times; in 
\citet{2009MNRAS.395.1146P} (hereafter P09) we explored how high mass 
X-ray binaries (HMXBs)\footnote{We assumed an initial primary mass of 
$\rm M_{\ast} \geq 8 \rm M_{\odot}$.} in primordial 
globular clusters at high redshifts could boost the UV ionizing power 
of the cluster. We found that harder ionizing spectra combined with
enhanced escape fractions for X-rays implied that HMXBs could be just
as efficient at ionizing the IGM as their MS progenitors. Similar 
ideas have been explored in \citet{2011A&A...528A.149M} and 
\citet{2011ApJ...738..163W}, in which the primary is a stellar mass 
black hole, as well as by \citet[][]{2012arXiv1206.1335M} in his 
census of potential sources of reionization.\\

The analysis in P09 focused on the ionizing power of HMXBs in 
globular clusters because (i) the inferred ages of metal poor globular 
clusters imply that they formed at $z \gtrsim 6$ 
\citep[cf.][]{2006ARA&A..44..193B}; (ii) the relationship between the 
initial mass functions (IMF) of stars and the dynamical evolution of 
clusters is well understood \citep[cf.][]{1997MNRAS.289..898V}; and 
(iii) the escape fraction for UV photons was likely to be large, 
assuming that globular clusters followed similar orbits then to the 
ones they follow now \citep[cf.][who explored the ionizing power of 
massive stars in globular clusters]{2002MNRAS.336L..33R}. This last 
point is important because the large UV escape fraction allowed P09 
to carry out a straightforward comparison of a globular cluster's UV 
and X-ray ionizing power.

However, as noted in P09, HMXBs are likely to be a generic 
by-product of high mass star formation \citep[cf.][]{2001ApJ...554...27H}, 
and should form as well in gas-rich galaxy discs as in young 
globular clusters. This is consistent with observations of star forming 
galaxies at $z \lesssim 1$ that show that luminous compact X-ray 
sources -- with properties mirroring those of HMXBs in our Galaxy -- 
are good indicators of recent star formation activity (see Figure 10 of 
\citealt{2012MNRAS.419.2095M} and Figure 2 of \citealt{2012arXiv1207.2157M}). 
In this paper, we extend the analysis presented in P09 to study 
HMXBs as generic sources of ionizing radiation. Here our focus is on refining our 
treatment of their spectral energy distribution. Rather than assuming a 
simple power-law form for the HMXB spectrum (as we did in P09), we use the spectrum
of Cygnus X-1, the archetypal Galactic HMXB \citep[e.g.][]{2006ARA&A..44...49R}, as 
our template. We assess how this impacts on the ionizing power of a coeval
stellar population over time and we quantify how the ionizing luminosity of the
population is boosted by the presence of HMXBs. Finally, we estimate the possible
contribution of HMXBs to the present-day soft X-ray background, assuming both our
new template spectrum and power-law spectra of the kind that have been used in 
previous studies (e.g. P09).\\

The structure of the remainder of this paper is as follows. In 
\S~\ref{sec:methods} we describe our time-dependent Monte Carlo model for the
spectral energy distribution of a coeval population of stars in which
HMXBs are forming. In \S~\ref{sec:results} we present results for the
time evolution of the ionizing power and spectral energy distribution 
of the population, and we quantify how HMXBs enhance the ionizing power
of the stellar population. In \S~\ref{sec:sxrb}, we show that the X-ray 
luminosity produced in our model does not violate observed limits on the 
soft X-ray background. Finally, we summarize our results in 
\S~\ref{sec:summary} and comment on their implications for cosmological 
reionization and high redshift galaxy formation.

\section{Methods}
\label{sec:methods}

\subsection{Modelling HMXBs in a Single Stellar Population}
\label{subsec:model} 

As in P09, we set up a Monte Carlo model of a stellar population, 
assumed to form in a single instantaneous burst, and follow the 
evolution of the massive stars over the first 250 million years, 
through their MS lives and into the HMXB phase. The main features of 
our model can be summarized as follows;

\paragraph*{The Massive Star Population:} We assume the IMF of 
\citet{2001MNRAS.322..231K} as our fiducial case, with stellar masses 
spanning the range $0.01 \leq {\rm M_{\ast}/M_{\odot}} \leq 100$, 
but we also verify our results for the \citet{1955ApJ...121..161S} and 
top-heavy \citet{2001ApJ...554.1274C} IMFs. All stars with 
$\rm M_{\ast} \geq 8 \rm M_{\odot}$ are assumed to form in binaries 
-- these are the progenitor population from which the HMXBs are 
drawn. Initial binary 
parameters are assigned following \cite{2006MNRAS.370.2079D} -- that is, 
companion masses are drawn from a uniform distribution between $0.01 
\leq {\rm M_{\ast}/M_{\odot}} \leq 100$ and orbital periods are 
distributed uniformly in logarithm between 1 and $10^4$ days. Massive 
star lifetimes are estimated using the results of 
\citet{2001A&A...371..152M,2003A&A...399..617M}, \cite{1993A&AS..102..339S} and 
\cite{2000A&A...361..101M} for metallicities of Z = 0, 0.008 and 0.02 
(i.e. solar metallicity) respectively; we explore the Z = 0 case in our results section.  

\paragraph*{The HMXB Population:} We assume that HMXBs form from 
binaries in which the initial MS mass of the primary exceeds $\rm 
M_{\ast} \sim 8 \rm M_{\odot}$, the threshold for neutron star formation 
\citep[cf. Figure 1 of][]{2003ApJ...591..288H}, and the donor (i.e. secondary) 
mass lies in the range $M_{\ast} \geq 3 \rm M_{\odot}$\footnote{This 
donor mass is lower than the usual definition of HMXBs -- donor OB stars 
\citep[cf. Table 1 of][]{2006ARA&A..44..323F} with typical masses 
$\gtrsim 10 \rm M_{\odot}$ \citep[e.g.][]{2012MNRAS.tmp.2855J}. However, it is
reasonable to include these intermediate mass X-ray binaries because they are
sufficiently luminous to contribute to the X-ray ionizing power of the stellar
population and their MS lifetimes are of order $\sim 10^8$ yrs.}. 
Once the primary goes supernova, we estimate the remnant mass using Figure 3 of 
\citet[][]{2003ApJ...591..288H}. The binary will be disrupted if it 
loses more than half its mass in the supernova -- which, for our 
fiducial Kroupa IMF, implies that approximately 30\% of binaries 
survive, thereby setting an upper limit to the total number of HMXBs 
that can potentially form. Following P09, we draw a survival fraction 
$f_{\rm sur}$ of this $\sim 30\%$ at random and consider them as HMXBs; 
$f_{\rm sur}$ captures the various uncertainties that prevent massive 
binaries from evolving into HMXBs. We assume that HMXBs are active until 
the companion star evolves off the main sequence and goes supernova. 

We have noted already that the binaries that survive are more likely to 
host black holes than neutron stars, especially in low metallicity 
systems where the formation rate of HMXBs in which the primary is a 
black hole could be a factor of $\sim$10 higher than at Solar 
metallicity \citep[cf.][]{2010ApJ...725.1984L,2012MNRAS.tmp.2855J}. 
There is also observational evidence that the black hole mass is likely 
to be larger in lower metallicity systems 
\citep[cf.][]{2010MNRAS.403L..41C}. This should, in principle, shape our 
HMXB luminosity function \citep[cf.][]{2006MNRAS.370.2079D}. However, 
we make the simplifying assumption to draw HMXB luminosities from a 
Weibull distribution with a peak luminosity of $L_{\rm X} \sim 
10^{38}$~erg~s$^{-1}$ but capped such that they do not accrete at 
super-Eddington rates; this sets an upper limit of approximately $L_{\rm 
X} \simeq 1.26 \times 10^{38} (\rm M/\rm M_{\odot})$~erg~s$^{-1}$ on the 
luminosity of an HMXB with primary mass $\rm M$. This approach gives a 
distribution that is consistent with the luminosities of compact X-ray 
sources in nearby galaxies whose X-ray binary populations are dominated 
by HMXBs \citep[cf. Figure 1 of ][]{2004NuPhS.132..369G}; see P09 for 
further discussion of this point.

\subsection{Spectral Energy Distribution \& Time Dependence}
\label{subsec:spectra} 

We split the spectral energy distribution of the stellar population 
into two components:

\paragraph*{Main Sequence (MS) Stars:} Each star is assumed to radiate as a black body 
with an effective temperature of

\begin{equation}
  \label{eq:teff}
  T_{\rm eff}=\left(\frac{L_{\ast}}{4\pi\,R_{\ast}^2\sigma_{\rm SB}}\right)^{1/4}
\end{equation}

\noindent where $R_{\ast}/R_{\odot}=(\rm M_{\ast}/\rm M_{\odot})^{0.8}$ 
is stellar radius and $\sigma_{\rm SB}$ is the Stefan-Boltzmann 
constant. Stellar luminosity $L_{\ast}$ is assumed to follow a mass-luminosity 
relation of the form

\begin{equation}
  \label{eq:mass-lum}
  \frac{L_{\ast}}{L_{\odot}}=\alpha\,\left(\frac{\rm M_{\ast}}{\rm M_{\odot}}\right)^{\beta},
\end{equation}

\noindent where $\alpha$, which governs the amplitude of the 
relation, and $\beta$, which governs its slope, are observed to 
depend on stellar mass \citep[e.g.][]{2004ASPC..318..159H}; their 
values and the stellar mass range in which they are applicable are 
summarized in Table~\ref{tab:bb_constants}. 

Empirically there is a wealth of evidence that the rate at which stellar 
luminosity varies with mass ($\beta=d\log L/d\log M$) decreases with increasing 
mass $\rm M_{\ast}$. For example, \citet{2007MNRAS.382.1073M} find 
$\beta \sim 4.1$ at $M \sim 1$~M$_{\odot}$ to $\beta \sim 3.2$ at 
$M \sim 20$~M$_{\odot}$ (see their Table 6), while \citet{2007AstL...33..251V} 
report that $\beta \sim 2.76$ over the mass range $20 \lesssim M/{\rm M}_{\odot} \lesssim 50$. 
Our values of $\alpha$ and $\beta$ provide a good approximation to the 
functional form presented in \citet{2007MNRAS.382.1073M} for
stellar masses $M_{\ast} \lesssim 50 M_{\odot}$. Above 
$50 M_{\odot}$ we must extrapolate because observationally inferred 
data are few; we assume that the relation is slightly shallower
than it is below $50 M_{\odot}$. This is an uncertainty, but it has 
negligible effect on our results because relatively few stars are formed 
in this mass range and their lifetimes are short. We find that the 
mass-luminosity relation can be well approximated by a $2^{\rm nd}$-order
polynomial of the form,
\begin{equation}
  \log L_{\ast}/L_{\odot}= c_0+c_1\log {\rm M_{\ast}}/{\rm M_{\odot}} + c_2 \left(\log {\rm M_{\ast}}/{\rm M_{\odot}}\right)^2.
\end{equation}
\noindent Here the coefficients have the values $c_0$=-0.04172, $c_1$=4.4954
and $c_2$=-0.6041, and $\log$ indicates logarithm base 10.

\begin{table}
 \centering
 \caption{Adopted Mass-Luminosity Relation.}
 \begin{tabular}{cccc}
   $M_{\rm min}/M_{\odot}$ & $M_{\rm max}/M_{\odot}$ & $\alpha$ & $\beta$\\
   \hline
   0.5          &   3          & 0.8    & 4.5   \\
   3            &   10         & 1.8    & 3.6  \\
   10           &   20         & 5.8    & 3.1  \\
   20           &   50         & 25.8   & 2.6  \\
   50           &   100        & 100.0    & 2.3  \\
   \hline
 \end{tabular}
 \label{tab:bb_constants}
\end{table}

\paragraph*{High Mass X-ray Binaries:} We model HMXB spectra using the 
archetypal Galactic HMXB Cygnus X-1 as our template. Cygnus X-1 is the 
brightest HMXB in the Galaxy and it has been studied in exquisite detail 
\citep[see, for example, the recent review by][]{2006ARA&A..44...49R}. It 
consists of a black hole of mass $8.7 {\pm} 0.8 \rm M_{\odot}$ \citep{2007ApJ...663..445S} 
and a super-giant companion (HDE 226868). Its spectrum is observed to 
fluctuate between distinct low-hard and high-soft states, examples of 
which can be found in \citet{1999MNRAS.309..496G}.

\begin{itemize}

\item The low-hard state is characterized by a luminosity of $L_{\rm 
2-10 keV} \sim$ $3{\times}10^{36}$~erg~s$^{-1}$ and a hard power-law 
spectrum ${\propto}E^{-{\Gamma}}$ with spectral index of ${\Gamma}$ 
${\sim}$ $1.6$-$1.8$ \citep[e.g ][]{1997MNRAS.288..958G,1999MNRAS.309..496G}.

\item The high-soft state is characterized by luminosity roughly one 
order of magnitude higher and a strong black body component with $kT$ 
${\sim}$ 0.5 keV with soft power-law tail with ${\Gamma}$ ${\sim}$ $2.5$ 
\citep[e.g.][]{1977Natur.267..813D,1982Natur.295..675O}.

\end{itemize}

\noindent This suggests that we should adopt distinct spectral shapes 
for low-hard and high-soft states. However, we understand neither what 
sets the duration of these two states -- for example, Cygnus X-1 is 
observed to be predominantly in its low hard state, but analogous 
systems such as LMC X-3 \citep{2007AIPC..924..530V} appear to spend more 
time in their high soft state -- nor how the nature and duration of 
these states depend on factors such as metallicity. For this reason, we 
make the simplifying assumption that HMXB spectra do not vary in time 
and instead introduce a threshold in X-ray luminosity of 
$10^{37}$~erg~s$^{-1}$, above (below) which a source is in a soft (hard) 
state. In particular, we model

\begin{itemize}

\item the low-hard state as a power-law of slope -0.8 between 
$2-10$~keV, the energy range in which this spectrum profile is observed, and

\item the high-soft state by a composite black body and power-law 
spectrum -- the power-law has a slope -1.1 and is normalized such that 
the ratio between the luminosities of the black body and power-law components 
matches that of Cygnus X-1, while the black body temperature is calculated by 
assuming ($L_{X}/L{_{\rm Cyg X-1}}$) = ($T/T{_{\rm Cyg X-1}}$)$^4$.

\end{itemize}

\section{Results}
\label{sec:results}

In the following subsections, we show how the ionizing power of a 
stellar population formed in an instantaneous burst evolves over the 
first 150 Myrs after formation (\S~\ref{ssec:ionizing_power}).
We compare and contrast results in the presence and absence of HMXBs 
and, where appropriate, we comment on the sensitivity of our results 
to the assumed value of the survival fraction ($f_{\rm sur}$) and our 
choice of IMF.

\subsection{Ionizing Power over Time}
\label{ssec:ionizing_power}

\begin{figure}
\centering
\plotone{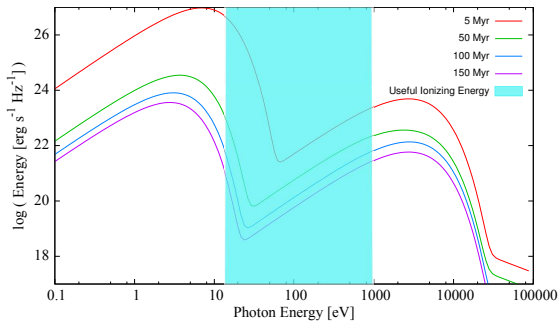}
\caption{Spectra of the total energy output of the fiducial stellar population
  (i.e. $N$=10$^6$, Kroupa IMF, $f_{\rm sur}=1$) after $5$, $50$, $100$, and
  $150$ Myrs (solid, long-, medium- and short-dashed curves respectively). The 
  spectral shape derives from the stellar UV black body and the HMXB softer 
  black body + harder power-law component, where we use the spectrum of Cygnus 
  X-1 for our HMXB template.}
\label{fig:time_evol}
\end{figure}

In Figure \ref{fig:time_evol} we show the spectral energy distribution 
of the fiducial stellar population (i.e. Kroupa IMF, $f_{\rm sur}=1$ and 
$N$=$10^6$) as a function of time -- a composite of MS stars, 
which we model individually as black bodies, and HMXBs, 
which we model as a combination of black bodies and power-law 
components. The amplitude of the black body corresponding to MS 
stars decreases with time, while its peak shifts to lower energies; 
this reflects the evolution of the most massive stars in the stellar 
population -- which dominate the UV-luminosity -- off the MS. Over 
the same period the amplitudes of both the black body and 
power-law components of the HMXBs decrease with time. 

The same qualitative trends can be seen if we vary the IMF from 
Kroupa to Chabrier or Salpeter -- there is a systematic increase 
(decrease) in the amplitude for the Chabrier (Salpeter) IMF, which 
reflects an increase (decrease) in the proportion of massive stars 
that form. If we vary the size of population between $N$=10$^4$ to 
$N$=10$^6$ the amplitude varies linearly with the size of 
population. Varying the survival fraction $f_{\rm sur}$ between 0.01 
and 1 suppresses the amplitude of the HMXB black body and power-
law contribution while leaving the black body contribution from the
MS unchanged. 

\begin{figure}
  \centering
 \includegraphics[height=0.65\textwidth,angle=-90]{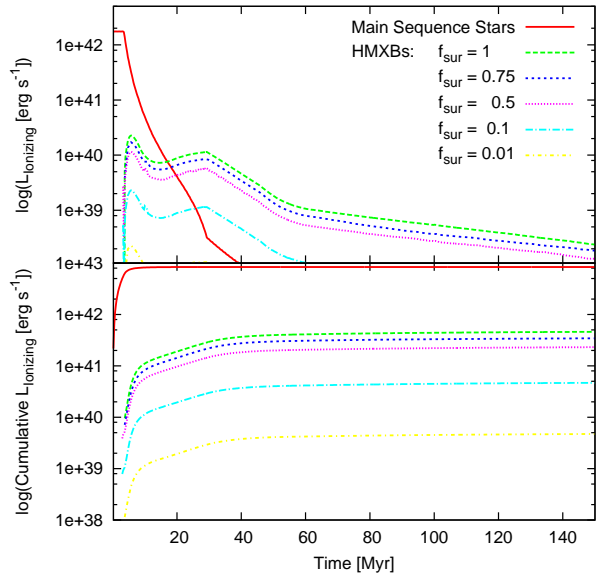}
 \caption{Time evolution of the instantaneous and cumulative ionizing power 
   of a stellar population of $N=10^6$ stars and a Kroupa IMF (upper and lower 
   panels respectively). Solid curves correspond to the contribution from MS 
   stars only; curves of different line types correspond to the 
   combined contribution of MS stars and HMXBs for five different values of 
   the survival fraction $f_{\rm sur}$ with $0.01 \leq f_{\rm sur} \leq 1$.}
  \label{fig:tot_energy}
\end{figure}

In Figure \ref{fig:tot_energy} we estimate how much of this energy
(instantaneous in the upper panel, cumulative in the lower panel) is 
available to ionize neutral atomic hydrogen (HI) in the IGM as a 
function of time. We evaluate this by integrating the spectra plotted in 
Figure \ref{fig:time_evol} between a lower limit of 13.6 eV, the 
minimum photon energy required to ionize atomic hydrogen, and an 
upper limit of $E_{\rm lim}$, the energy above which atomic hydrogen 
becomes transparent to hard X-rays\footnote{As in P09, we estimate 
$E_{\rm lim}$ by requiring that $\sigma(E_{\rm lim})=c^{-1}H(z)\,n_{\rm H}(z)^{-1}$ 
where $\sigma(E)$ is the ionization cross section of neutral hydrogen, 
$H(z)$ is the Hubble parameter at redshift $z$, $n_{\rm H}(z)$ is the mean 
number density of hydrogen and $c$ is the speed of light.}; for the redshift range 
we consider ($z \gtrsim 6$), $E_{\rm lim} \gtrsim 1$ keV. Provided 
$f_{\rm sur}>0.1$, we find that HMXBs dominate the ionizing power 
of the stellar population after $\sim 20$ Myrs. 

Interestingly we note that the X-ray ionizing power shows a 
pronounced bump between $20 \lesssim t \lesssim 40$ Myrs. This is an imprint 
of our definition of an HMXB as a binary system in which the mass of the 
primary must exceed $\sim 8 \rm M_{\odot}$ at the end of its MS lifetime  -- 
the threshold for neutron star formation. Because the MS lifetime of an 
$8 \rm M_{\odot}$ is $\sim 25$ Myrs 
\citep[cf.][]{2001A&A...371..152M,2003A&A...399..617M} and because 
statistically we expect more HMXBs to be drawn from stars close to this 
$8 \rm M_{\odot}$ limit, we expect an excess in HMXBs to form around this 
time and to produce a bump in the ionizing power. After this time, HMXBs can 
no longer form and the ionizing power comes from a population that is in 
terminal decline.

\begin{figure}
  \centering
  \includegraphics[width=0.5\textwidth, trim=0cm 3.5cm 0cm 0cm, clip=false]{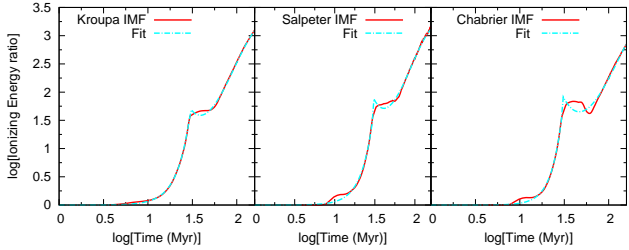}
  \caption{Time evolution of the boost factor 
  $f_{\rm HMXB}=L_{\rm HMXB}/L_{\rm MS}$ for a stellar population 
  of $N=10^6$ stars, a survival fraction $f_{\rm sur} = 1$ and a 
  Kroupa, Salpeter and Chabrier IMFs (left, middle and right panels
  respectively). Dotted-dashed curves correspond to polynomial
  fits of the form given by Equation~\ref{eq:boost} with 
  parameters given in Table~\ref{tab:params}.}
  \label{fig:fits}
\end{figure}

We quantify how HMXBs enhance the ionizing power of the stellar 
population by measuring the boost factor 
$f_{\rm HMXB}=L_{\rm HMXB}/L_{\rm MS}$, which compares the
ionizing luminosity of the MS stars only ($L_{\rm MS}$) and of the 
combined MS stars and HMXBs ($L_{\rm HMXB}$). In 
Figure~\ref{fig:fits} we show how $f_{\rm HMXB}$ varies 
with time for a stellar population of $10^6$ stars and 
$f_{\rm sur}$=1 assuming Kroupa, Salpeter and Chabrier IMFs 
respectively (solid curves, left to right panels). Because the ionizing 
power of the MS component declines sharply after $\sim 10-20$ Myrs 
as the most massive stars end their lives, $f_{\rm HMXB}$ grows 
steadily with time, peaking at $f_{\rm HMXB} \sim 1600$ after 
$t \sim 180$ Myrs before dropping off abruptly. The dashed curves in 
Figure~\ref{fig:fits} show fits to the numerical results
that agree to better than 10\%. We fit the results with 2 formulae, 
below and above the bump at 30 Myr. These are given by

\begin{eqnarray}
  \label{eq:boost}
  \log(f_{\rm HMXB}) =  \left\{ \begin{array}{ll}
  10^{-5}\text{ }p\text{ }e^{\text{ }q\log{t}} &  t < 30\\
  a+b\log{t}+c(\log{t})^2 & \\
  \hspace{2cm}+d(\log{t})^3 & t > 30\\
  \end{array}	\right.
\end{eqnarray}
where $t$ is population age in Myr, and are 
valid up to approximately 180 Myr, above which HMXB contribution falls 
rapidly to 0 (and $f_{\rm HMXB}$ drops to 1). The parameters $p$, $q$ and 
$a$ to $d$ are tabulated in Table~\ref{tab:params}.

\begin{table}
 \centering
 \caption{Boost Factor $f_{\rm HMXB}$ Parameters.}
 \begin{tabular}{lcccccc}
   IMF & $p$ & $q$ & $a$ & $b$ & $c$ & $d$\\
   \hline
   Kroupa  & 7.85 & 6.73 & 54.2  & -85.4 & 45.0 & -7.61\\
   Salpeter& 6.53 & 6.82 & 66.2 & -104 & 55.1 & -9.40 \\
   Chabrier& 0.852 & 8.23 & 46.8 & -69.2 & 34.3 & -5.42 \\
   \hline
 \end{tabular}
 \label{tab:params}
\end{table}

\section{Cosmic Soft X-ray Background}
\label{sec:sxrb}

The Universe at $z \gtrsim 6$ is effectively transparent to photons with 
energies $\gtrsim 1$ keV. This means that hard photons remain unabsorbed 
by neutral hydrogen and so a hard X-ray background was built up in the
early Universe that, through redshifting, contributes to the present day
soft X-ray background (SXRB), the mean specific background 
intensity of soft X-rays. We wish to determine the contribution made 
to the SXRB by the HMXB population present at $z \gtrsim 6$ and whether 
or not their contribution violates observed limits. Equivalent calculations 
have been performed by \citet{2004ApJ...613..646D}, 
\citet{2005MNRAS.362L..50S} and \citet{2007MNRAS.375.1269Z} 
for mini-quasars in the early Universe. However, we adopt the 
approach of \citet{2012MNRAS.421..213D}, who looked at the contribution 
of star-forming galaxies to both the soft and hard  X-ray backgrounds from 
high redshift to the present day.\\

In what follows, we assume values of $H_0=100h \,\rm km\,s^{-1}\,Mpc^{-1}$
for the Hubble parameter with $h$=0.71, $\Omega_{\rm M}=0.266$ for the matter density 
parameter, and $\Omega_{\Lambda}=0.734$ for the dark energy density parameter 
\citep[cf.][]{2011ApJS..192...18K}, and we take as our observed limit on the SXRB 
the value of $\sim 3.4 \pm 1.4 \times 10^{-13} {\rm erg}\,{\rm s}^{-1}{\rm cm}^{-1}
{\rm deg}^{-2}$ obtained by \citet{2007ApJ...671.1523H} for the excess flux between 
$1-2$~keV, which corresponds to hard X-rays at $z \gtrsim 6$. We estimate the 
contribution made by high redshift HMXBs to the observed SXRB by evaluating

\begin{equation}
  {\rm SXRB} = \frac{\Delta\Omega}{4\pi}\frac{c}{H_0}\int_6^{z_{\rm max}}\frac{\dot{\rho}_{\ast}\mathcal{L}_{\rm X}(z)}{(1+z)^2\mathcal{E}(z)} dz,
\label{eq:sxrb}
\end{equation}

\noindent where $\Delta\Omega\sim 3 \times 10^{-4}\,{\rm sr}\,{\rm deg}^{-2}$; 
$c$ is the speed of light; $z_{\rm max}$ is the earliest (unknown) redshift at which HMXB 
formation proceeds; $\mathcal{E}(z)=\sqrt{\Omega_{\rm M}(1+z)^3+\Omega_{\Lambda}}$; 
$\dot{\rho}_{\ast}$ is the comoving star formation rate density in units of 
$\rm M_{\odot}{\rm yr}^{-1}{\rm Mpc}^{-3}$; and $\mathcal{L}_{\rm X}(z)$ 
is the ``K-corrected'' X-ray luminosity per unit star formation rate in the
observed energy range $X=E_1-E_2$.

For $\dot{\rho}_{\ast}$ we explore four possibilities, spanning the range of
potential star formation histories. We take (i) a fixed value of 
$\dot{\rho}_{\ast}(z=6) \simeq 0.002$~M$_{\odot}$~yr$^{-1}$~Mpc$^{-1}$, 
as measured at $z$=6 by \citet[][]{2010MNRAS.409..855B}; (ii) a fixed value of 
$\dot{\rho}_{\ast}(z=6) \simeq 0.17$~M$_{\odot}$~yr$^{-1}$~Mpc$^{-1}$, as 
measured at $z$=6 by \citet{2012ApJ...754...46T}; (iii) a fixed value of 
$\dot{\rho}_{\ast}(z=6) \simeq 0.02$~M$_{\odot}$~yr$^{-1}$~Mpc$^{-1}$, estimated 
at $z$=6 using the functional form of \citet{2006ApJ...651..142H},

\begin{equation}
  {\dot{\rho}_{\ast}}(z) = \frac{(u+vz)h}{1+(z/w)^x},
  \label{eq:rhostar}
\end{equation}

\noindent with parameters $u$=0.017, $v$=0.13, $w$=3.3 and $x$=5.3 from 
\citet{2001MNRAS.326..255C}; and (iv) a redshift-dependent value of $\dot{\rho}_{\ast}$, 
estimated using Equation~\ref{eq:rhostar}.

\begin{figure}
  \centering
  \plotone{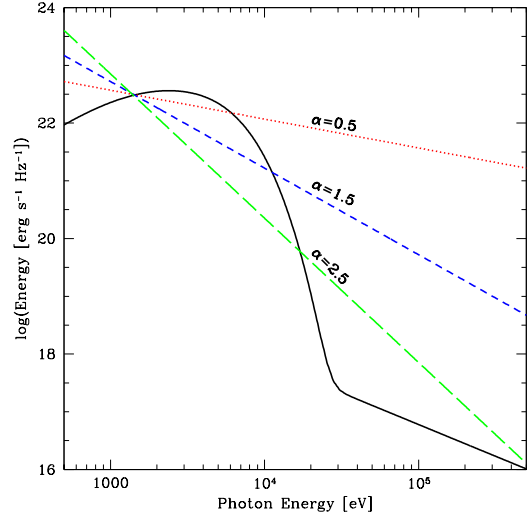}
  \caption{Comparison of the composite (i.e. black-body plus power-law) HMXB
    spectral energy distribution (solid curve) and possible power-law 
    alternatives (with indices $\alpha$=0.5,1.5,2.5, corresponding to dotted, 
    dashed and dotted-dashed curves respectively). The composite HMXB spectrum
    is as measured at $t$=50 Myrs in Figure~\ref{fig:time_evol}, while the
    power-law spectra are normalized such that the total energy emitted between
    1 and 2 keV is the same for both the composite and power-law spectra.}
  \label{fig:compare_spectra}
\end{figure}

For $\mathcal{L}_{\rm X}(z)$ we follow \citet{2012MNRAS.421..213D} and 
recast it as $\mathcal{L}_{\rm X}(z)=c_{\rm X}K(z)$. Here we assume the range of 
values for 
$2.6 \leq c_{\rm X}/10^{39} {\rm erg}{\rm s}^{-1}{(\rm M_{\odot} yr^{-1})^{-1}} \leq 3.7$
measured by \citet{2012arXiv1207.2157M} for compact resolved X-ray sources in galaxies
(lower limit) and unresolved galaxies in the Chandra Deep Field North and 
ultra-luminous infra-red galaxies (upper limit); and we calculate 
\begin{equation}
  K_{}(z) = \frac{\int_{E_1(1+z)}^{E_2(1+z)} EF(E)dE}{\int_{0.5\rm keV(1+z)}^{8\rm keV(1+z)} EF(E)dE}
  \label{eq:kz}
\end{equation}
\noindent where $F(E)$ represents the spectrum of a single HMXB, $E_1$=1 keV 
and $E_2$=2 keV. Examples of the spectra we consider are shown in Figure~\ref{fig:compare_spectra} -- the solid curve corresponds to the cumulative spectrum
derived from HMXBs in the fiducial stellar population described in \S3, as 
measured at $t$=50 Myrs\footnote{We treat the unnormalized cumulative spectrum
as indicative of the shape of a typical HMXB.}, while the dashed, dotted and 
dotted-dashed curves correspond to power-law spectra ($\propto E^{-\alpha}$) 
that have been normalized such that the total energy emitted between 1 and 2 
keV is the same for both composite and power-law spectra. The units are 
arbitrary -- it is the shape, not the amplitude, that is important when 
evaluating Equation~\ref{eq:kz} -- and it is worth noting how poorly 
simple power-law spectra approximate the shape of the more realistic composite
spectrum.

In all cases we fix our lower bound in Equation~\ref{eq:sxrb} at $z=6$ and 
allow our upper bound $z_{\rm max}$ to vary between $z=20$ and $z$=50. This corresponds to a 
range in age of the Universe of between 48 and $\sim$ 180 Myrs old ($z$=50 and 20 respectively), 
which is likely to bracket the redshifts when the first generation of dark matter halos 
were sufficiently massive to support cooling and subsequent star formation formed 
\citep[cf. Figure 1 of ][]{2005SSRv..117..445G}. However, we note that our results 
are insensitive to $z_{\rm max}$ for the range of values that we consider.\\

Evaluating Equation~(\ref{eq:sxrb}), we find that the 
contribution of HMXBs to the soft X-ray background is of order 
$\sim 5 \times 10^{-16}$ erg s$^{-1}$ cm$^{-2}$ deg$^{-2}$, depending on what 
precisely is assumed for the star formation rate density and spectrum --
roughly a factor of $\sim$ 1000 smaller than the observed limit of 
$\sim 3.4 \times 10^{-13}$ erg s$^{-1}$ cm$^{-2}$ deg$^{-2}$ measured by
\citet{2007ApJ...671.1523H}. In general, the higher the star formation rate 
density, the larger the contribution that is
possible, because of our assumption that HMXB formation tracks (massive) star 
formation. If power-law spectra are adopted, we find that SXRB $\sim 2.15 \times 10^{-16}$ erg s$^{-1}$ cm$^{-2}$ deg$^{-2}$ for $\alpha$=0 compared to 
$\sim 11.4 \times 10^{-16}$ erg s$^{-1}$ cm$^{-2}$ deg$^{-2}$ for $\alpha$=2; that
is, the softer the spectrum, the larger the contribution HMXBs can make. 
Nevertheless, this contribution is still a factor of $\sim 100$ smaller than 
the observed limit.

\section{Summary}
\label{sec:summary}

There are compelling astrophysical reasons to expect that HMXBs, 
which are observed to be a natural by-product of massive star 
formation \citep[e.g.][]{2012MNRAS.419.2095M,2012MNRAS.421..213D}, could 
be an important source of feedback in galaxy formation over cosmic 
time \citep[e.g.][]{2012MNRAS.tmp.2855J}. Because the cross section 
for photons to be absorbed by neutral hydrogen decreases strongly 
with energy as $E_{\gamma}^{-3}$, X-rays from HMXBs can effectively diffuse 
through the IGM and deposit their energy at relatively large distances 
from the stellar population, in an act of non-local feedback \citep[a similar
effect is expected for mini-quasars; see, for example,][]{2005MNRAS.360L..64Z}. 
This contrasts with lower energy UV photons, whose correspondingly larger 
absorption cross section means that they ionize the IGM in the vicinity of 
the stellar population, in an act of local feedback.

In this paper, we build on the work of \citet{2009MNRAS.395.1146P} in 
which we examined the X-ray luminosity and effective ionizing 
power\footnote{We made the simplifying assumption that an 
energetic X-ray photon can ionize multiple hydrogen atoms, and so treated 
secondary electrons that ionize hydrogen atoms as effective photons.}
of a coeval population of stars and HMXBs at $z \gtrsim 6$. 
\begin{itemize}
\item We have used the archetypal Galactic HMXB, Cygnus X-1, as our template for a more
realistic HMXB spectrum. The form of this spectrum depends on the accretion
state of the system and is a composite of black-body and power-law components. 
\item Using this new composite spectrum, we have quantified how HMXBs enhance the ionizing 
power of the stellar population by defining the boost factor 
$f_{\rm HMXB}=L_{\rm HMXB}/L_{\rm MS}$ (where $L_{\rm MS}$ and $L_{\rm HMXB}$ are the 
ionizing luminosities of a population of MS stars and a combined population of 
MS stars and HMXBs respectively) and characterizing and its variation with time. 
Because the ionizing power of the MS component declines sharply 
after $\sim$ 10-20 Myrs as the most massive stars end their lives, we 
find that $f_{\rm HMXB}$ grows steadily with time before peaking at 
$t \sim 100$ Myrs and after dropping off abruptly. This 
demonstrates the relatively long-lived nature of HMXBs as ionizing sources
when compared to young massive stars on the MS.
\item Finally, we have estimated the contribution of HMXBs to the soft X-ray 
background by assuming that HMXB formation is linked explicitly 
to the global star formation rate density, and we show that it does not 
violate observed limits. Depending on what is assumed for the star formation
rate density, we estimate a contribution of $\sim 5 \times 10^{-16}$ erg 
s$^{-1}$ cm$^{-2}$ deg$^{-2}$ if HMXB spectra are modelled as composite black-bodies
and power-laws  -- roughly a factor of $\sim$ 1000 smaller than 
the observed limit of $\sim 3.4 \times 10^{-13}$ erg s$^{-1}$ cm$^{-2}$ deg$^{-2}$ 
measured by \citet{2007ApJ...671.1523H}.
\end{itemize}

We do not compare the HMXB contribution 
to the ionizing X-ray background present during reionization to
the contribution from other source populations, such as mini-quasars
and accreting super-massive black holes, in this paper. Such a 
comparison has been carried out by \citet{2012arXiv1206.1335M}
who concludes that HMXBs were likely to be a minor contributor to 
reionization. This may be the case -- although the conclusion 
depends on the assumed spectral properties of the various sources 
considered, which in most cases are uncertain -- but our work 
implies that HMXBs can have a significant impact on 
the ionization structure and heating rate of the IGM at $z \gtrsim 6$,
extending the volumes of partially ionized hydrogen by 
factors of $\gtrsim$ 1000. This will have a direct impact on the efficiency 
of galaxy formation by, for example, suppressing the collapse of gas onto low-mass 
dark matter halos \citep[e.g.][]{2003MNRAS.338..273M,2012arXiv1205.6467T}  
and modifying the cooling rate of hot gas in galaxy halos 
\citep[e.g.][]{2010MNRAS.403L..16C}. We will demonstrate this explicitly 
in a forthcoming paper (James et al., in preparation), in which we will 
use the more realistic composite spectrum in a radiative
transfer calculation to assess the relative importance of such HMXB 
feedback in heating the high redshift IGM and the potential HMXB contribution 
to both cosmological reionization and galaxy formation in general.

\section*{Acknowledgements}

We thank the anonymous referee for their insightful report that has helped
improve this paper. CP, GJ, CC and GW acknowledge the support of the 
theoretical astrophysics STFC rolling grant at the University of Leicester. 
Part of the research presented in this paper was undertaken as part of the 
Survey Simulation Pipeline 
(SSimPL; {\texttt{http://www.astronomy.swin.edu.au/SSimPL/}). The 
Centre for All-Sky Astrophysics is an Australian Research Council 
Centre of Excellence, funded by grant CE11E0090.

\vspace{0.1cm}

\bibliographystyle{apj}

\end{document}